\documentclass[prb,reprint]{revtex4-2} 

\usepackage{amsmath}  
\usepackage{amsfonts} 
\usepackage{graphicx} 
\usepackage{tabularx}
\usepackage{array}
\usepackage{booktabs}
\usepackage{siunitx}
\usepackage{multirow}
\usepackage{float}

\begin{document}

\title{Why and how to implement worked examples in upper division theoretical physics}

\author{Philipp Scheiger}
\email{philipp.scheiger@uni-jena.de}
\author{Holger Cartarius}
\affiliation{Research Group Teaching Methodology in Physics and Astronomy, Friedrich-Schiller-University Jena, 07743 Jena}

\author{Ronny Nawrodt}
\affiliation{Physics Didactics Research, 5. Physikalisches Institut and Center for Integrated Quantum Science and Technology, Universität Stuttgart, Pfaffenwaldring 57, Stuttgart 70569, Germany}

\begin{abstract}
Studying worked examples has been shown by extensive research to be an effective method for learning to solve well-structured problems in physics and mathematics. The effectiveness of learning with worked examples has been demonstrated and documented in many research projects. In this work, we propose a new four-step approach for teaching with worked examples that includes writing explanations and finding and correcting errors.  This teaching method can even be implemented in courses in which homework performance constitutes part of the grading system.  This four-step approach is illustrated in the context of Lagrangian mechanics, which is ideal for the application of worked examples due to its universal approach to solve problems.
\end{abstract}

\maketitle

\section{Introduction}
Worked examples are step-by-step solutions of exercises or tasks.
When well-structured worked examples are included in the learning process of novices, they improve the initial acquisition of cognitive skills as compared to conventional problem-solving activities (the so-called worked-example effect) \cite{Sweller.1985, Paas.2006, Sweller.2011}. Effective learning with worked examples can consist, for example, of studying several solutions and explaining them oneself before solving a similar problem without additional support (example-problem pairs \cite{Sweller.1985}).

The advantages of this method are often understood through cognitive load theory \cite{Sweller.2011}. This theory assumes that learning is associated with cognitive load and describes what can facilitate or impede learning. For example, the removal of extraneous tasks such as remembering how to complete mathematical operations allows the student to concentrate on the main learning task. In addition, worked examples are considered cognitive activating, meaning that they cause the student to focus attention on the most important learning tasks (for an overview, see Refs. \cite{Sweller.1998, Renkl.2002, Sweller.2011}). 
Learning with worked examples can be more effective than to conventional problem solving because when learners are presented with a new topic, they often concentrate cognitive resources on the technical aspects of solving the problem. In that case, less resources are available for the construction of abstract schemata, which lead to transfer and can help to solve related problems (cf.\ Ref.\ \cite{Renkl.2003}). 
The effectiveness of such a worked examples effect is well examined and proven in well-structured areas of mathematics and physics \cite{Sweller.2011, Cooper.1987, Paas.1992, Paas.1994, Ward.1990, vanGog.2011, Elby.2000}, and also in text comprehension \cite{Oksa.2010}, and essay writing \cite{Kyun.2013}.

However, worked examples are not always superior or more efficient in learning in comparison to classical problem solving \cite{Kalyuga.2001}. They need to be structured and designed in a way that extraneous load is decreased and germane load is increased such that learners profit from them. For example, self-explanations by students turn out to be important \cite{Renkl.1997}. The request for self-explanations (and also finding or fixing errors \cite{Groe.2007, Brown.2016}) in our approach is an important difference from the simple presentation of solved problems in lectures. 

The added value of worked examples has also been demonstrated for double integrals in calculus at university level \cite{Santosa.2019}, i.e.,\ in a topic which is needed regularly in theoretical physics. Therefore, worked examples are a promising tool for improving teaching in upper-division theoretical physics. They can be used as further methodical variation (in addition to, e.g., peer instruction and small group tutorials) for the transformation of upper-division physics courses \cite{Chasteen.2015, Chasteen.2008, Goldhaber.2009, Pollock.2012} or to help students to overcome typical difficulties with mathematical tools \cite{Wilcox.2013, Caballero.2015, Pepper.2012}.

In addition to example-problem pairs (the simplest and most often recommended scheme of worked examples \cite{Sweller.1985,vanGog.2011}), other approaches with additional intermediate tasks can be beneficial \cite{Atkinson.2003} such as including self-explanation prompts and successively removing more and more worked-out tasks in solutions. It is the purpose of this work to show that a four-step approach based on worked examples can indeed be introduced in theoretical physics. In Sec.\ \ref{sec:concept} we present the concept and explain its
introduction in exercise courses. Short notes on our experiences from several courses are given in Sec.\ \ref{sec:experience}, and conclusions are drawn in
Sec.\ \ref{sec:conclusion}. More details based on the relevant effects from cognitive load theory (cf.\ Refs. \cite{Trafton.1993, Sweller.2011, Rodiawati.2019, Chi.1989, Renkl.1997, Hausmann.2002, Schworm.2006, RobertS.Siegler.2008, Durkin.2012, Sweller.1985,  vanGog.2011, Groe.2007, Stark.2011, Atkinson.2003, Paas.2003}) relevant for the concept and our conclusions for worked examples in theoretical physics can be found in the supplementary material bellow this paper. 

\section{Concept of a four-step approach applied in theoretical physics}
\label{sec:concept}

The goal of the four-step approach is to strengthen the worked example effect. It ensures that extended self-explanations are taken seriously by all students due to two additional steps embedded between completely elaborated worked examples and problem tasks without solutions. While the individual steps are known and tested, we propose an arrangement ideally suited for courses that assign problem sets. A lecturer can implement the method with very little additional preparation time because it can be based on existing textbook problems. It offers the students a broader variety of examples without investing more time, and the opportunity to acquire a deeper understanding of the underlying principles used in the exercises. The concept is structured as follows.

The first step in our scheme is a maximally elaborated worked example with a detailed step-by-step solution. The problem chosen for this step should be paradigmatic for this type of physics task. To reduce extraneous load generated by the necessary collection of information (split attention effect \cite{Sweller.2011}) the important aspects are highlighted in the problem description and the structure of the solution is explicitly shown. Ideally the calculation path and the explanation of the different solution tasks are close together on the worksheet. The students' mathematical capabilities should determine amount of mathematical detail that is included.

After this preparation, we foster self-explanations in step two by demanding written explanations from the students for a solution that is presented without explanations. This example can be simpler than the first one to help students get started. Since most learners are passive or superficial self-explainers, \cite{Renkl.1997} we recommend either an obligatory written explanation for assignments that are submitted for grading or peer discussions for assignments that are completed in tutorials.

In step three students are asked not only to explain the work, but also to identify and correct errors that are purposefully included in the solution. This task should be more challenging than step 2; therefore, the chosen example should be more difficult again. We recommend 2-4 errors, such that the students continue to search after finding the first error but the wrong solution does not become too confusing. The focus should be on errors in the translation of the physics problem into its mathematical description. One calculation error should be included for a quick boost of accomplishment (see supplementary material).

Step four consists of a problem without any solution and it is the student's task to develop and provide a full solution, which is the typical last step of worked examples.

A detailed elaborate example for Lagrangian mechanics can be found in supplementary material 2 (bellow this text), which fits in a 90-min exercise class or can be completed by students at home. All examples are typical textbook problems \cite{Nolting.2011, Bartelmann.2018} (an English version of the first book is also available \cite{Nolting.2016}) and can be adopted easily. In addition, we provide a discussion about the lessons we have learned using worked examples, how they can be embedded in instruction, and their applicability in other topics.

\section{Experiences with the four-step approach}
\label{sec:experience}

We used this or a similar structure of worked examples in a set of different courses from mathematics refresher courses to seminars accompanying classical mechanics, electrodynamics, quantum theory, and thermodynamics/statistis lectures. In all these setups we observed strong interest of the students and their active participation. In each of our applications the majority of the students was able to solve the last problem correctly. Thus, we found that worked examples can be productive for several teaching situations in theoretical physics. The example presented in supplementary material 2 (see bellow this text) on Lagrangian mechanics was tested in a 90-min online seminar with a very satisfying performance of the students. Some students were able to complete the whole task within the seminar; a short homework (completing the last example) remained for the others. 

While our four-step approach is, in principle, very simple, we want to emphasize the challenges. A great challenge is finding of the correct level of difficulty for the learners. Novices with less prior knowledge need more explanations, and error finding can overstrain them. However, redundant information can reduce the learning effects for students with adequate prior knowledge. Those two sides on the scale should be considered for every lecture and course.

\section{Conclusions}
\label{sec:conclusion}

Due to our positive experiences, we recommend the application of worked examples in theoretical physics and invite other lecturers to adopt and enhance the approach of worked examples presented here in theoretical physics.

A nice side effect of worked examples with additional tasks (written self-explanations and error finding) is that they can also be used to assess the students' performances. Suggestions on how this can be implemented can be found in supplementary material 2.

Our presented scheme can be applied to every topic for which a universal solution structure to problems exists and worked examples are usable in general. Very typical examples are Lagrangian mechanics of the second kind and solutions to the time-independent Schrödinger equation (see Table \ref{solutions_tasks}). 

\begin{table*}[t]
  \centering
  \caption{Solution tasks in Langrangian mechanics of the second type and the time-independent Schrödinger equation}
  \begin{ruledtabular}
    \begin{tabular}{p{1.5cm} p{7.0cm} p{0.5cm} p{7.0cm}}
      &Lagrangian mechanics &  &Time-independent Schrödinger equation \\
      \hline
      Task 1 &Set up the holonomic constrains and calculate the degrees of freedom $S$. &  &Determine the dimension of the Hilbert space and set up the Hamiltonian according to the physics problem. \\
      Task 2 &Define generalized coordinates $q_i$ according to the holonomic constrains. &  &Set up the eigenvalue equation. \\
      Task 3 &Set up the kinetic $T$ and potential $V$ energy. &  &Determine the eigenvalues.  \\
      Task 4 &Set up the Lagrangian. &  &Determine the eigenvectors according to their eigenvalues. \\
      Task 5 &Determine the equations of motion for every generalized coordinate. &  &Check for boundary conditions.  \\
      Task 6 &Reduce the equations if possible. & &  \\
    \end{tabular}
  \end{ruledtabular}
  \label{solutions_tasks}
\end{table*}

A systematic evaluation of the long-term performance of the students after attending the worked-examples problem courses has to be relegated to future research. This was not a topic in this work; however, an actual proof of the superiority as compared to bare problem solving is desirable. Nevertheless, previous research indicates a benefit of worked examples in university environments \cite{Santosa.2019}.

\section*{AUTHOR DECLARATIONS}
The authors have no conflicts to disclose.

\begin{acknowledgments}
	
	The authors are grateful for financial support by the Academy of Teaching Support of Friedrich Schiller University Jena. This research was also funded by the Bundesministerium für Bildung und Forschung (Federal Ministry of Education and Research), at the Professional School of Education Stuttgart Ludwigsburg in the project ``Lehrerbildung PLUS,'' grant No. 01JA1907A. This project is part of the ``Qualitätsoffensive Lehrerbildung,'' a joint initiative of the Federal Government and the Länder which aims to improve the quality of teacher training.
	The authors also thank Thomas Rubitzko for stimulating discussions at the beginning of this project.
	
\end{acknowledgments}

\appendix
\section{THEORETICAL FRAMEWORK} 
As highlighted in the main text, worked examples need to be well structured and embedded in the learning process of novices to improve the initial skill acquisition \cite{Sweller.1985, Paas.2006, Sweller.2011}.
The superiority of learning with worked examples as compared to conventional problem solving is often explained by the cognitive load theory (e.g., \cite{Paas.2003, Sweller.1998}). Cognitive load can be described as the amount of working memory resources a person needs to fulfill a task. The cognitive load theory describes three types of cognitive load that are relevant during learning. There is the intrinsic load which refers to the complexity of the learning contents in relation to a learner's prior knowledge. The second type, the extraneous load, refers to activities that are irrelevant (in general or under certain circumstances) for learning. For example, the search for new or forgotten information, such as the definition of a formula or a calculation rule, generates extraneous load if it is not a particular requested learning goal. In well designed worked examples this type of cognitive load is reduced as much as possible. The third type, called germane load, refers to cognitive resources that are bound by learning-relevant activities. Unlike intrinsic cognitive load, which is generally considered invariant, teachers can also manipulate germane load. Increased germane load aims not only to solve the task (intrinsic load) but to maximize the learning effect in automating schema and building understanding. Self-explanations of the step-by-step solution is an example for such cognitive load. In contrast to extraneous load germane load is desired and crucial when deep understanding of the learning contents is requested.

The worked examples effect can be effective in learning because when learners solve problems at the beginning of an unknown topic, significant cognitive resources are invested in technical aspects the solution, rather than dedicated to the construction of abstract schemata. It's this construction that leads to transfer and can help to solve related problems. Solving problems at the beginning is usually accomplished by a means-ends-analysis strategy. Here a substantial portion of cognitive load is needed to keep many aspects in mind such as the current problem state, the goal state, differences between them, etc. Remaining working memory capacities are therefore limited and are no longer available for learning processes like the construction of abstract schemata, which are crucial in theoretical physics. In contrast, learners can concentrate on gaining understanding when studying worked examples since they are freed from performance demands (cf. Ref. \cite{Renkl.2003}).

Cognitive load theory explains why learning with worked examples is superior to common problem solving in a broad variety of learning situations. So far there few examples of applications in upper division physics. However, there are no reasons why there should not be a positive effect. In particular, examples from mathematics show that even complex tasks can successfully be addressed with worked examples \cite{Santosa.2019}. 
This renders them to a promising tool for improving teaching in upper division theoretical physics. Thus, worked examples should be useful in the transformation of upper-division physics courses as it is pursued, e.g., in the Science Education Initiative (for an overview see \cite{Chasteen.2015, Chasteen.2008, Goldhaber.2009, Pollock.2012}). In addition, worked examples might help to support students to overcome typical difficulties with certain mathematical tools (see, e.g. \cite{Wilcox.2013, Caballero.2015, Pepper.2012}) when they have to solve problems with long and complex mathematical transformations.

\subsection{DESINGS FOR WORKED EXAMPLES} \label{WE}
Worked examples are not always superior or more efficient in learning as compared to problem solving \cite{Kalyuga.2001}. However when worked examples are structured and designed in a way that extraneous load is decreased and germane load is increased, learners do profit from them.

A good design to reduce extraneous load takes several effects of the cognitive load theory into account (cf. \cite{Trafton.1993, Sweller.2011, Rodiawati.2019}). These are the split-attention effect, the redundancy effect and the expertise reversal effect. When learners have to split their attention between at least two sources of information that have been separated either spatially or temporally, extraneous cognitive load is increased and the learning efficiency is decreased (split attention effect). Information that is redundant or unnecessary has the same effect on learners cognitive load (redundancy effect) or information that is already known by learners and stored in their long-term memory (expertise reversal effect). In order to motivate learners to study the examples and to process the information in worked examples it is recommended to demand them to solve a similar problem after studying the worked example. 

Besides decreasing extraneous cognitive load we want to increase germane load. Therefore we use the self-explanation effect and the studying error principle. We want to foster self explanation because learners who study examples longer and explain them more actively to themselves are more successful \cite{Chi.1989}. But most learners are passive or superficial self-explainers according to Renkl \cite{Renkl.1997}. This leads to the conclusion that demanding self-explanations by instructional procedures is crucial in learning from worked examples. Self-explanation does not necessarily mean a talking-aloud procedure. There is evidence that prompting written self-explanations foster learning outcomes \cite{Hausmann.2002, Schworm.2006}. The self-explanation effect can be increased by explaining correct and incorrect solutions \cite{RobertS.Siegler.2008}; especially explaining why incorrect solutions are wrong helps to avoid these errors later \cite{Durkin.2012}. The task to correct their own mistakes in mid-term exams can help students to increase their performance in the final exam \cite{Brown.2016}. However there is only a positive effect in finding and explaining errors in worked examples for learners with adequate prior knowledge \cite{Groe.2007}, so implementing errors in worked examples too early can overwhelm weaker learners with little prior knowledge. To prohibit such an overchallenge, weaker learners need additional support by explicitly marking errors \cite{Groe.2007} or by expert explanations and feedback why certain steps in the solution are correct or incorrect \cite{Stark.2011}.

Next to the most recommended scheme of worked examples, viz.\ example-problem pairs \cite{Sweller.1985,vanGog.2011} other approaches with more intermediate steps are possible. In many tasks it can be beneficial to combine self-explanation prompts and fading or successively removing more and more worked-out steps in solutions as Atkinson et al.\ \cite{Atkinson.2003} have shown. They found that this combination even positively influences the quality of example processing for far-transfer tasks, thus a scenario in which we wish to apply worked examples.

\subsection{CONCLUSIONS FOR THE DESIGN IN THEORETICAL PHYSICS}\label{Implementation}
To be most efficient in learning scenarios, worked examples must be designed according to the effects and principles described above. However the more complex the problem is, the more complex it is to take account of all these effects in the given solution. This relation leads to a special challenge in theoretical physics since it requires a broad spectrum of previous knowledge and skills (mathematically and physically). However, this does not mean that worked examples are not applicable. Santosa \cite{Santosa.2019} implied that worked examples can increase the learning efficiency in complex tasks of calculus such as higher-dimensional integrals, exactly the type of mathematics many physics problems demand to get solved. This result provides a strong motivation to extend this approach to exercises in theoretical physics.

In addition to physical principles (e.g. laws, boundary conditions, etc.), mathematical challenges (such as differential equations, differential \& integral calculus, linear algebra and so on) also increase the difficulty of physics tasks. When students are not well versed in mathematical techniques they have to interrupt their thoughts on the intended problem with a search for (mathematical) solution methods. A search that can lead to other textbooks or lectures as sources of information creates extraneous load due to the split-attention effect. 
Thus the split attention effect is most crucial in the applications we have in mind to reduce extraneous cognitive load. Reducing the spatial split attention effect in such complex exercises is challenging due to the amount of information. Nevertheless it can be minimized to a certain degree by highlighting important information, correlations or dependencies in the problem. In addition offering a step-by-step solution of a worked example should reduce the temporal split attention effect.

Since students of theoretical physics are not total novices in problem solving, our conclusion is mainly to foster self-explanations to increase germane cognitive load in the continuative examples. One possibility is to request written explanations for certain or all steps in the solution from students \cite{Schworm.2006, Hausmann.2002}. Another possibility is to extend the examples by certain comprehension questions students are requested to answer. When students have shown adequate problem solving skills or a broad prior knowledge, finding and correcting errors should improve the effects of self-explanation even more.

\section{FOUR-STEP APPROACH IN LAGRANGIAN MECHANICS}\label{Lagrange}
This supplement provides a paradigm of how worked examples could be implemented in theoretical physics. Worked examples are useful in every topic in which there is a universal solution structure. To illustrate this approach, we have chosen the topic of Lagrangian mechanics of the second kind. The solution structure behind various application examples is itself a prime example of solution structures and therefore predestined to illustrate ``worked examples.''
\begin{enumerate}
	\item Define the holonomic constraints and calculate the degrees of freedom $S$.
	\item Define generalized coordinates $q_i$ according to the holonomic constrains.
	\item Write expressions for the kinetic $T$ and potential $V$ energy.
	\item Set up the Lagrangian.
	\item Determine the equations of motion for every generalized coordinate.
	\item Reduce the equations if possible.
\end{enumerate}
In addition, in theoretical physics, Lagrangian mechanics is usually the first exposure of university students to a new topic, which is more conceptual than Newtonian mechanics, and thus additional support is appropriate here. The examples used in this paradigm can be found in modified form in textbooks like \cite{Nolting.2011, Bartelmann.2018}. An English version of the first book is also available \cite{Nolting.2016}. 

The first step in our scheme is a fully worked example with a detailed step-by-step solution. After this preparation we foster self-explanations in step two by demanding written explanations from the students to accompany a solution that is provided. In step three this is extended by finding and fixing errors with an explanation. Step four consists of a problem without any solution and it is the student's task to develop and provide a full solution. 

The whole program fits in a 90 minute presence exercise course or can be dealt with by students at home.  All examples are typical text book problems and can be adopted easily. 

\subsection{Step 1: Maximally elaborate solution - Pendulum on springs}

As mentioned in the main text, the first step in our scheme is a fully worked example with a detailed step-by-step solution. The split attention effect is reduced by highlighting the important aspects in the problem description. Explanations are given right beside each task of the solution procedure. Mathematical explanations should be given where necessary depending on the mathematical capabilities of the learners.

\begin{figure}[]
	\centering
	\includegraphics[width=0.33\textwidth, angle=0]{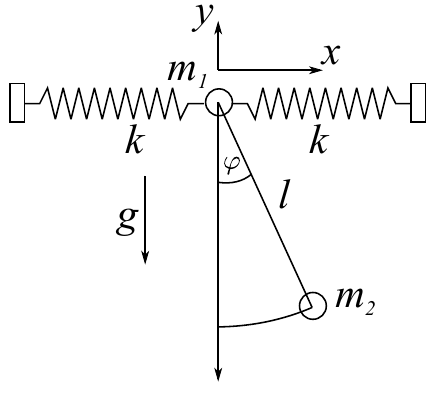}
	\caption{Sketch of the two masses of example 1. $m_1$ is clamped between two springs and $m_2$ is a pendulum fixed to $m_1$ in a gravitational field.}
	\label{img:Lagrange1}
\end{figure}

In the context of Lagrangian mechanics, the pendulum on springs (see Fig. \ref{img:Lagrange1}) was chosen as first example since it contains many typical features of textbook exercises in Lagrangian mechanics. Thus, the detailed solutions of this problem offers the learners a broad overview. Students can learn how to deal with two coupled masses and two types of potential energy. For novices in Lagrangian mechanics we recommend a very detailed solution for example by explicitly naming all dimensions and arguing via holonomic constraints the irrelevance of the $z$ component. If the students are already familiar with holonomic constraints this part is redundant and should be excluded from the worked example.

\textbf{Exercise:} Set up the Lagrangian (or Lagrangian function) and determine the equations of motion. Reduce them as much as possible.

A point mass $m_1$ is attached via \underline{two springs with the same spring constant $k$} between two walls. The equilibrium point of the springs corresponds to the center position of the point mass between those walls. \underline{The restoring force is $|\mathbf{F}| = k |\mathbf{x}|$}. Mass $m_1$ is only able to move horizontally along the $x$-axis. A second point mass $m_2$ is attached to $m_1$ via \underline{a massless rod with length $l$}. The second point mass can oscillate in the $x$-$y$-plane under influence of a homogeneous gravitational field with \underline{force $\mathbf{F_G} = -m_2 g \mathbf{e_y}$}. The angle of deflection is given by $\varphi$ (see Fig. \ref{img:Lagrange1}). The small-angle approximation applies. 

Table \ref{tab:step1} shows the solution path alongside detailed explanations for each calculation step. 

\begin{table*}[bth]
	\centering
	\caption{Worked example for step 1. Next to the solution steps are detailed explanations to support the students.}\label{tab:step1}
	\begin{tabular*}{\textwidth}{p{5cm} p{12.5cm}}
		\hline
		\hline
		\multicolumn{2}{l}{\textbf{Set up the holonomic constrains and calculate the degrees of freedom $S$.}}\\
		{$\!\begin{aligned} 
				z_1 = 0, \; &z_2 = 0 \\
				y_1 = 0, \;	&(x_1-x_2)^2+y_2^2 = l^2 
			\end{aligned}$}
		&\vspace{-0.5cm} The masses do not move in the $z_1$, $z_2$ and $y_1$ coordinate. 
		The $x_2$, $y_2$ components of mass 2 depend on $x_1$ and the rod's length $l$. The dependency is defined by the Pythagorean theorem.\\
		{$\!\begin{aligned} 
				S=6-4=2
			\end{aligned}$}
		&Two free particles have six degrees of freedom (2 times the 3 spatial dimensions). Since there are four constraints, the total number of degrees of freedom reduces down to two.\\
		\hline
		\multicolumn{2}{l}{\textbf{Define generalized coordinates $q_i$ according to the holonomic constrains.}}\\
		{$\!\begin{aligned} 
				q_1 = x_1=x, \qquad &q_2 = \varphi 
			\end{aligned}$}
		& A wise choice for the generalized coordinates is the x position of mass 1 and the angle of deflection.\\
		\vspace{-0.5cm}
		{$\!\begin{aligned} 
				x_1 = x, \qquad &x_2 = x + l \sin \varphi \\
				y_1 = 0, \qquad &y_2 = - l \cos \varphi \\
				z_1 = 0, \qquad &z_2 = 0 
			\end{aligned}$} 
		& The coordinates are chosen such that mass $m_1$ is in the center in the equilibrium state. 
		The transformation between Cartesian and the generalized coordinates is helpful for setting up the Lagrangian.\\
		{$\!\begin{aligned}
				\dot{x}_1 = \dot{x}, \qquad &\dot{x}_2 = \dot{x} + l \dot{\varphi} \cos \varphi \\
				\dot{y}_1 = 0, \qquad &\dot{y}_2 = l \dot{\varphi} \sin \varphi \\
				\dot{z}_1 = 0, \qquad &\dot{z}_2 = 0
			\end{aligned}$}  
		& \vspace{-0.5cm} The derivative of time delivers the required velocity. Therefore the time dependency of all variables and the chain rule must be taken into account. 
	\end{tabular*}
\begin{tabular*}{\textwidth}{p{8cm} p{9.5cm}}
	\hline
	\multicolumn{2}{l}{\textbf{Set up the kinetic $T$ and potential $V$ energy.}}\\
	{$\!\begin{aligned} 
			T &= \frac{m_1}{2} \dot{x}^2_1 + \frac{m_2}{2} \left(\dot{x}^2_2 +  \dot{y}^2_2 \right) \\
			&= \frac{m_1}{2} \dot{x}^2 + \frac{m_2}{2}(\dot{x}^2 + l^2 \dot{\varphi}^2  + 2l\dot{x}\dot{\varphi} \cos \varphi )
		\end{aligned}$}
	& \vspace{-0.7cm} The overall kinetic energy is the sum of all velocity components. $y_1$, $z_1$ and $z_2$ contribute nothing since there is no motion in these directions.\\
	{$\!\begin{aligned} 
			V_1 &= \frac{k}{2} (-x)^2 + \frac{k}{2} x^2 =  kx^2  \\
			V_2 &= -m_2 g l \cos \varphi 
		\end{aligned}$}
	& \vspace{-0.5cm} The potential energy of the first point mass is the energy stored in the two springs. The
	second mass swings in a homogeneous gravitational field.\\
		\hline
		\multicolumn{2}{l}{\textbf{Set up the Lagrangian.}}\\
		{$\!\begin{aligned} 
				L &= T-V  \\
				&= \frac{m_1 + m_2}{2} \dot{x}^2 + \frac{m_2}{2} l^2 \dot{\varphi}^2 + m_2 l (\dot{x} \dot{\varphi} + g) \cos \varphi -kx^2
			\end{aligned}$}\vspace{0.1cm}
		& \vspace{-0.6cm}The Lagrangian is the overall kinetic energy minus the potential energy.\\
		\hline
		\multicolumn{2}{l}{\textbf{Determine the equations of motion for every generalized coordinate.}}\\
		{$\!\begin{aligned} 
				&\frac{d}{dt} \frac{\partial L}{\partial \dot{x}} - \frac{\partial L}{\partial x} = 0 \\
				& = \frac{d}{dt} \left[(m_1 + m_2) \dot{x} + m_2 l \dot{\varphi} \cos \varphi \right] + 2kx\\
				& = (m_1 + m_2) \ddot{x} + m_2 l \ddot{\varphi} \cos \varphi - m_2 l \dot{\varphi}^2 \sin \varphi + 2kx
			\end{aligned}$}
		& \vspace{-1.0cm} Via the Euler-Lagrange equations the equations of motion can be determined. To do so, the derivatives must be calculated in the correct order.\\
		&\\
		{$\!\begin{aligned} 
				&\frac{d}{dt} \frac{\partial L}{\partial \dot{\varphi}} - \frac{\partial L}{\partial \varphi} = 0 \\
				&= \frac{d}{dt} \left[m_2 l^2 \dot{\varphi} + m_2 l \dot{x} \cos \varphi \right] + m_2 l (\dot{x} \dot{\varphi} + g) \sin \varphi \\
				&= m_2 l \ddot{\varphi} + m_2 \ddot{x} \cos \varphi + m_2 g \sin \varphi
			\end{aligned}$}
		&\\
		\hline		
		\multicolumn{2}{l}{\textbf{Reduce the equations if possible.}}\\
		{$\!\begin{aligned}
				(m_1 +m_2) \ddot{x} + 2kx &= m_2 l (\dot{\varphi}^2 \varphi-\ddot{\varphi})\\
				&\\
				l \ddot{\varphi} + g \varphi &= -\ddot{x}
			\end{aligned}$}
		& \vspace{-0.8cm}The approximation of small angles ($\sin \varphi \approx \varphi$, $\cos \varphi \approx 1$) simplifies expansions of trigonometric functions and reduces the equations of motion.\\
		\multicolumn{2}{p{15cm}}{These two differential equations are strongly coupled and as the term $m_2l\dot{\varphi}^2$ indicates not even linearly coupled but quadratically. It is not trivial to solve such equations, but it is also not part of this exercise.}\\
		\hline
		\hline
	\end{tabular*}
\end{table*}

\newpage
\subsection{Step 2: Fostering self-explanations - Mass in a cone in a gravitational field}

In this example, students are encouraged to formulate their own explanations. A simpler example is sufficient for this task since the self-explanation is in the focus. This helps the students to get started. 

\begin{figure}[]
	\centering
	\includegraphics[width=0.285\textwidth, angle=0]{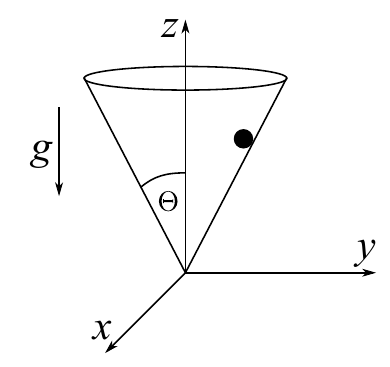}
	\caption{Sketch of example 2. A pellet moves under influence of gravity on the inner surface of a cone with an aperture angle $\Theta$.}
	\label{img:Lagrange2}
\end{figure}

In Lagrangian mechanics the problem ``Mass in a cone in a gravitational field'' (see Fig. \ref{img:Lagrange2} was chosen as second example. The problem is not as difficult as example 1, since there is only one mass and one potential energy.

\textbf{Exercise:} Set up the Lagrangian and determine the equations of motion. Reduce them as much as possible.

On the inner face of an upwards opened cone is a pellet with mass $m$. The cone has an aperture angle of $2\Theta$ and the pellet can move frictionless on its inner surface. It is under the influence of the gravitational force $\mathbf{F_G} = - mg \mathbf{e_z}$. The axis of the cone is identical with the $z$-axis and its top is located in the origin, see Fig. \ref{img:Lagrange2}.

Table \ref{tab:step2} shows the mathematical solution.

\begin{table*}[bth]
	\centering
	\caption{Mathematical solution for step 2. Next to the solution steps is free space for students' written self-explanations.}\label{tab:step2}
	\begin{ruledtabular}
	\begin{tabular*}{\textwidth}{p{7.25cm} p{8.75cm}}
		\multicolumn{2}{l}{\textbf{Set up the holonomic constrains and calculate the degrees of freedom $S$}}\\
		{$\!\begin{aligned} 
				\tan \Theta = \frac{\rho}{z} \quad \Leftrightarrow \quad z = \rho \cot \Theta \\
				S = 3-1 = 2
			\end{aligned}$}
		& \\ 
		\hline
		\multicolumn{2}{l}{\textbf{Define generalized coordinates $q_i$ accordant to the holonomic constrains.}} \\
		{$\!\begin{aligned} 
				q_1 &= \rho, \qquad q_2 = \varphi \\	
				x &= \rho \cos \varphi, \; y = \rho \sin \varphi,\; z = \rho \cot \Theta 
			\end{aligned}$}
		& \\ 
		\hline
		\multicolumn{2}{l}{\textbf{Set up the kinetic $T$ and potential $V$ energy.}}\\
		{$\!\begin{aligned} 
				T &= \frac{m}{2}[\dot{x}^2 + \dot{y}^2 +\dot{z}^2] \\
				&= \frac{m}{2} \left[ (1+ \cot^2 \Theta) \dot{\rho}^2 + \rho^2 \dot{\varphi}^2 \right] \\
				V &= mgz = mg \rho \cot \Theta
			\end{aligned}$}
		& \\
		\hline
		\multicolumn{2}{l}{\textbf{Set up the Lagrangian.}}\\
		{$\!\begin{aligned} 
				L &= T-V  \\
				&= \frac{m}{2} \left[ (1+ \cot^2 \Theta) \dot{\rho}^2 + \rho^2 \dot{\varphi}^2 \right]  \\
				&\qquad - mg \rho \cot \Theta
			\end{aligned}$}
		& \\
		\hline
		\multicolumn{2}{l}{\textbf{Determine the equations of motion for every generalized coordinate.}}\\
		{$\!\begin{aligned} 
				&\frac{d}{dt} \frac{\partial L}{\partial \dot{\rho}} - \frac{\partial L}{\partial \rho} = 0 \\
				&m(1+ \cot^2 \Theta) \ddot{\rho} - m(\rho \dot{\varphi}^2 - g \cot \Theta) = 0
			\end{aligned}$}
		& \\
		{$\!\begin{aligned} 
				&\frac{d}{dt} \frac{\partial L}{\partial \dot{\varphi}} - \frac{\partial L}{\partial \varphi} = 0 \\
				& m(\rho^2 \ddot{\varphi} + 2\rho \dot{\rho} \dot{\varphi}) = 0 
			\end{aligned}$}
		&\\
		\hline
		\multicolumn{2}{l}{\textbf{Reduce the equations if possible.}}\\
		{$\!\begin{aligned}
				(1+ \cot^2 \Theta) \ddot{\rho} - \rho \dot{\varphi}^2 + g \cot \Theta = 0\\
				\rho^2 \ddot{\varphi} + 2\rho \dot{\rho} \dot{\varphi} = 0, \qquad \rho \neq 0 
			\end{aligned}$} 
		& \\
	\end{tabular*}
\end{ruledtabular}
\end{table*}

\subsection{Step 3: Finding and fixing errors - Two masses on a wedge coupled by a spring}
\begin{figure}[t]
	\centering
	\includegraphics[width=0.33\textwidth, angle=0]{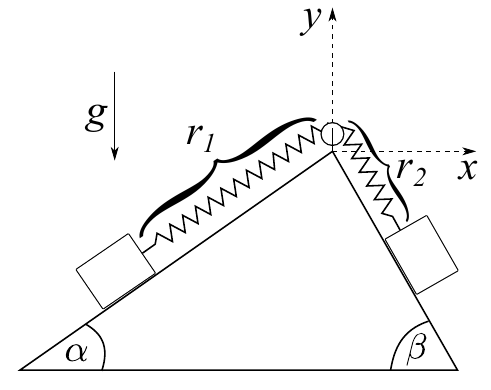}
	\caption{Sketch of example 3. Two masses on a wedge connected via a massless spring in a gravitational field.}
	\label{img:Lagrange3}
\end{figure}

To foster self-explanation the third example has errors intentionally included. This task should be more challenging than step 2; therefore the chosen example should be more difficult. In our example for Lagrangian mechanics, an example with two generalized coordinates (Fig. \ref{img:Lagrange3}) was chosen such that the errors cannot be found easily. In the incorrect solution a holonomic constraint is wrong. This error was chosen because students often have problems with the transition between mathematics and physics. The second error is located in the potential energy of the spring. Only the extension or compression compered to the stress-free length is relevant for the potential energy. The missing stress-free length in the potential is a typical student mistake and is easily overlooked in a given solution. 

\textbf{Exercise:} Set up the Lagrangian and determine the equations of motion of the problem given in Fig. \ref{img:Lagrange3}.

Two masses $m_1$ and $m_2$ move on a wedge. There is no friction but a gravitational field given by the force $\mathbf{F_G} = -m_{1/2} g \mathbf{e_y} $. The masses are connected via a massless spring with a spring rate $k$ and a stress-free length $l$.\\

\textbf{Set up the holonomic constrains and determine the number $S$ of degrees of freedom}
\begin{align*}
	z_1 = 0, \qquad z_2 = 0, \qquad	\frac{y_1}{x_1} = \tan \alpha , \qquad \frac{y_2}{x_2} = \tan \beta
\end{align*}
\begin{equation*}
	S = 6 - 4 = 2
\end{equation*}
\textbf{Define generalized coordinates, which conform to the holonomic constrains}
\begin{equation*}
	q_1 = r_1, \qquad q_2=r_2
\end{equation*}
\textbf{Transformations}
\begin{align*}
	x_1 &= -r_1 \cos \alpha, \qquad &x_2 = r_2\cos \beta \\
	y_1 &= -r_1 \sin \alpha, \qquad &y_2 = -r_2 \sin \beta \\
	z_1 &= 0, \qquad \qquad \quad &z_2 = 0
\end{align*}
\textbf{Set up the kinetic $T$ and potential $V$ energy}
\begin{align*}
	T &= \frac{m_1}{2}\dot{r}_1^2 (\cos^2 \alpha +\sin^2 \alpha) + \frac{m_2}{2} \dot{r}_2^2 (\cos^2 \beta +\sin^2 \beta) \\
	&= \frac{1}{2} (m_1 \dot{r}_1^2 + m_2\dot{r}_2^2)
\end{align*}
\begin{equation*}
	V = -m_1 g r_1 \sin \alpha - m_2 g r_2 \sin \beta + \frac{k}{2} (r_1+r_2)^2
\end{equation*}
\textbf{Set up the Lagrangian}
\begin{align*}
	L = T-V = &\frac{1}{2} (m_1 \dot{r}_1^2 + m_2\dot{r}_2^2) + m_1 g r_1 \sin \alpha \\
	&\qquad + m_2 g r_2 \sin \beta - \frac{k}{2} (r_1+r_2)^2
\end{align*}
\textbf{Determine the equations of motion for every variable}

\begin{minipage}[]{0.49\textwidth}
	\begin{align*}
		\frac{d}{dt} \frac{\partial L}{\partial \dot{r}_1} - \frac{\partial L}{\partial r_1} = 0 \\
		m_1 \ddot{r}_1 - m_1 g \sin \alpha + k(r_1 + r_2) = 0
	\end{align*}
\end{minipage}
\begin{minipage}[]{0.49\textwidth}
	\begin{align*}
		\frac{d}{dt} \frac{\partial L}{\partial \dot{r}_2} - \frac{\partial L}{\partial r_2} = 0 \\
		m_2 \ddot{r}_2 - m_2 g \sin \beta + k(r_1 + r_2) = 0
	\end{align*}
\end{minipage}

\subsection{Step 4: Student's task - Rotating mass on a tabletop connected to a second hanging mass}
\begin{figure}[b]
	\centering
	\includegraphics[width=0.33\textwidth, angle=0]{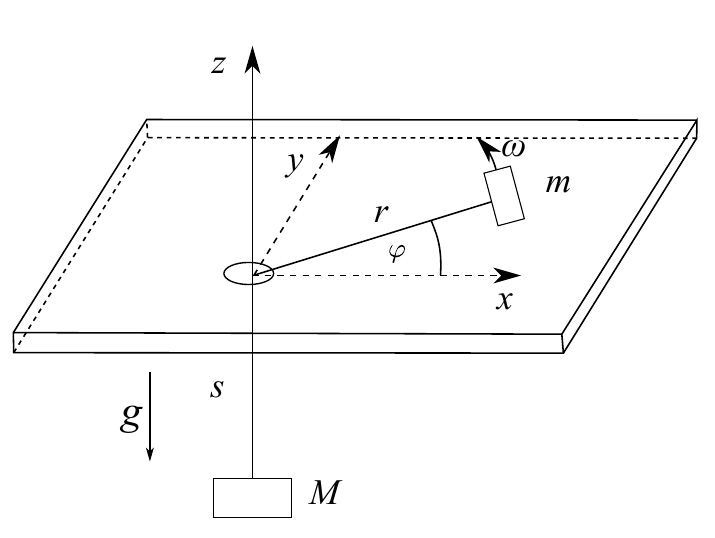}
	\caption{Sketch of example 4. A rotating mass on a tabletop is pulled towards an hole by a second mass in a gravitational field.}
	\label{img:Lagrange4}
\end{figure}

Since learners take the self explanation more seriously, when they have to solve a familiar problem by their own, the last step is the most important. Here a suitable task contains again all typical elements but is not too easy or is a transfer task. 
In Lagrangian mechanics we chose (Fig. \ref{img:Lagrange4}) a rotating mass on a tabletop connected to a second hanging mass as a fourth example since many elements of the first three examples can be used in this solution. The coupling of two masses can be found in examples 1 \& 3, and the rotary motion in example 2. However, the combination of both effects never appeared. Thus, there is a transfer students have to master solving this problem.

\textbf{Exercise: } Set up the Lagrangian and determine the equations of motion of the following setup. Reduce them as much as possible.
A mass $m$ rotates without friction on a tabletop. The mass is connected to a second mass $M$ by a string of the length $l=r+s$. The second mass is below the tabletop and is pulled downwards by the influence of the gravitational force $\mathbf{F_G}=-Mg\mathbf{e_z}$.

\subsection{Discussion}\label{Discussion}

\subsubsection{Lessons learned}
As mentioned in the main text, we used this or a similar structure of worked examples in two mathematics refresher courses prior to theoretical physics lectures in classical mechanics and electrodynamics, in a further topic of classical mechanics (Noether theorem), in electrodynamics (profiles for Maxwell's equations and derivation of the wave equation), in quantum theory (verbalization of formulas) and thermodynamics/statistical physics (idealized cycles and partition functions). In all these setups we observed active participation of our students. The worked example approach was preferred by students in initial test phases to the presentation of solutions by the lecturer and was rated as more useful. The performance of our students in the last problem was satisfying. We were also able to show our students more problems in each topic this way. Thus, we experienced that worked examples can be productive for several teaching situations in theoretical physics.  

While the scheme presented here is in principle very simple, we want to mention two aspects, which should be kept in mind. The first addresses the prior knowledge of the learners. The great challenge for lecturers, who want to use worked examples, is to adapt the correct level of difficulty for the learners. Novices need more solution steps and explanations. Also error finding can overstrain them. On the other hand redundant information can reduce the learning effects for students with adequate prior knowledge. 
For example, in previous tests of worked examples in Lagrangian mechanics of the first kind we delivered only a mathematical solution without explanations. Since our students were not total novices, we thought additional explanation were redundant. But this approach overstrained them. Thus, our recommendation is to deliver more rather than fewer explanations in the beginning. 

For step 3 (finding and fixing errors) we recommend 2 to 4 errors, so that the students continue to search after the first finding but the wrong solution does not become too confusing. The type of errors should depend on the learning objectives of the instructor. We typically focus on physically incorrect terms but also miscalculations are possible to increase the number of errors. In general we recommend errors in the translation of the physics problem into its mathematical description, one calculation error for a quick boost of accomplishment, and errors that are typical for the used kind of problems. Here it can be helpful to skip the third step in one year and to search in students solutions for patterns in their mistakes.

In addition, when existing problem sets are used for the worked examples, the notion should be the same in every example, otherwise extraneous cognitive load can be increased and the worked examples lose their efficiency. If these two aspects are considered, we are convinced worked examples improve lectures and exercises in theoretical physics as an upper division levels research in mathematics \cite{Santosa.2019} and our own experience implicate.

\subsubsection{Embedding in instruction}
A beneficial side effect of the self-explanation fostering variants are their use for lecture certificates or ungraded semester performances. In some physics study programs students have to solve a certain amount of problems every
week to obtain an admission to the exam. This practice is justified with the training students get by solving problems as a preparation for the exam and it is meant to be a protection for students not to take an exam il-prepared.
As mentioned above worked examples can be superior to problem solving. This is why we suggest to exchange the often used problem solving exercises by worked examples which demand self-explanations or finding and fixing errors.

However, if the students do not solve all problems by themselves and a rating of their work is required, the rating system has to be adapted. Instead of taking just the number of correct solution steps or solved problems into account, it is possible to establish a rating on the steps of the worked example scheme presented above. The quality of the students explanations in the self-explanation tasks as well as the number of detected errors in the error finding problem can legitimate the lecture certificate. In particular, the number of problems that can be utilized for the rating does not change much. We have had good experience by exchanging two problems without assistance with four worked examples of the types described above.

The scheme presented does not need to be rigidly adopted by other instructors. Most important are the presentation of the solution structure (step 1) and the student's task (step 4). The steps in between can help to ease the transition from understanding a solution to producing a solution yourself. Therefore, this four-step approach can also be split or modified and adapted to time and structural constraints. While the four-steps approach filled 90 minutes in our case, the four steps can be split up due to time restrictions into consecutive seminars (down to 45 minute seminars). The time constraints and the level of the students must be weighed here. It is conceivable to outsource steps 1 and 2 in the homework or to omit one step. For teaching students we once skipped step 2 in favor of a educational introduction of worked examples and cognitive load theory.

\subsubsection{Applicability of the structure in other topics}
In every topic were there is a universal solution structure to problems similar to that in Lagrangian mechanics our presented scheme and worked examples in general are applicable. For example solving the time-independent Schrödinger equation
\begin{equation}
	\operatorname{\hat H}|\Psi\rangle = E |\Psi\rangle
\end{equation}
with a solution structure like:
\begin{enumerate}
	\item Determine the dimension of the Hilbert space and set up the Hamiltonian according to the physics problem.
	\item Set up the eigenvalue equation.
	\item Determine the eigenvalues.
	\item Determine the eigenvectors according to their eigenvalues.
	\item Check for boundary conditions. 
\end{enumerate}
Since physics problems occur in well structured domains, there are many topics suitable for worked examples. Just to inspire the readers mind we want to mention a few examples among many others. There are calculations with work integrals, or inertia tensors, Noether's theorem and Hamiltonian mechanics in classical mechanics. In quantum physics besides the time-independent Schrödinger equation there are perturbation theory and the determination of Clebsch-Gordan coefficients. In classical electromagnetism many solution methods of Poisson's equation like the multipole expansion are suitable for worked examples as well as are calculations for thermodynamic cycles or the determination of partition functions in statistical mechanics. 

\end{document}